\documentclass{article}
\usepackage[utf8]{inputenc}

\usepackage[scale=0.75]{geometry}

\usepackage{float}
\usepackage{amsfonts}
\usepackage{mathtools}
\usepackage{authblk}

\title{Structured Mean-field Variational Inference and Learning in Winner-take-all Spiking Neural Networks}

\author[1]{Shashwat Shukla}
\author[2]{Hideaki Shimazaki}
\author[1]{Udayan Ganguly}
\affil[1]{Indian Institute of Technology Bombay, India}
\affil[2]{Graduate School of Informatics, Kyoto University, Japan}

\date{\vspace{-3ex}}

\begin{document}
	
	\maketitle

\begin{abstract}
The Bayesian view of the brain hypothesizes that the brain constructs a generative model of the world, and uses it to make inferences following Bayes' rule. Although many types of approximate inference schemes have been proposed for hierarchical Bayesian models of the brain, the questions of how these distinct inference procedures can be realized by hierarchical networks of spiking neurons remains largely unresolved. Based on a previously proposed multi-compartment neuron model in which dendrites perform logarithmic compression, and stochastic spiking winner-take-all (WTA) circuits in which firing probability of each neuron is normalized by activities of other neurons, here we construct Spiking Neural Networks that perform \emph{structured} mean-field variational inference and learning, on hierarchical directed probabilistic graphical models with discrete random variables. In these models, we do away with symmetric synaptic weights previously assumed for \emph{unstructured} mean-field variational inference by learning both the feedback and feedforward weights separately. The resulting online learning rules take the form of an error-modulated local Spike-Timing-Dependent Plasticity rule. 
Importantly, we consider two types of WTA circuits in which only one neuron is allowed to fire at a time (hard WTA) or neurons can fire independently (soft WTA), which makes neurons in these circuits operate in regimes of temporal and rate coding respectively. We show how the hard WTA circuits can be used to perform Gibbs sampling whereas the soft WTA circuits can be used to implement a message passing algorithm that computes the marginals approximately. Notably, a simple change in the amount of lateral inhibition realizes switching between the hard and soft WTA spiking regimes. Hence the proposed network provides a unified view of the two previously disparate modes of inference and coding by spiking neurons.
\end{abstract}

\section{Introduction}

The Bayesian brain hypothesis posits that the brain encodes a generative model of its sensorium and causal hidden states using probability distributions, and uses this model to effectively incorporate uncertainty in its computations underlying action and perception \cite{doya2007bayesian,knill2004bayesian}. From this perspective, perception pertains to inverting this generative model by computing the posterior over hidden states given sensory input. The marginal of the sensory states with respect to the generative model appears as the denominator in the Bayes rule expression for the posterior. However, computing this marginal becomes computationally intractable given the large state-space of the hidden states. Furthermore, to learn the generative model under the maximum likelihood principle, we need to compute gradients of the model parameters with respect to this computationally intractable marginal. Variational inference is a method of computing an approximate posterior by converting this into an optimization problem \cite{blei2017variational,zhang2018advances}. The approximate posterior is chosen from a parametrized family of probability distributions and these parameters are optimized to minimize the Kullback-Leibler (KL)-divergence between the exact and approximate posterior. This procedure minimizes the variational free energy. Learning is achieved by maximizing an upper bound on the log marginal probability of observed sensory states, with respect to the parameters of the generative model. Since the variational free-energy is negative of the upper bound, learning corresponds to minimizing the free-energy. Hence by performing gradient descent for the parameters of the variational posterior as well as generative model, with the variational free energy as a cost function, we can learn the generative model as well as obtain an approximate posterior for it \cite{shimazaki2019principles}. 

The Free Energy Principle (FEP) claims that the brain is minimizing such a variational free-energy, and this principle has emerged as a unifying framework for modeling brain function \cite{friston2010free}. While FEP can potentially explain many observed neural phenomena through modeling at the systems level, we wish to address the question of how such schemes can be realized using biologically plausible neural circuitry described at the level of spikes. Pecevski et al. \cite{pecevski2011probabilistic} demonstrated how a spiking winner-take-all (WTA) circuit can perform exact inference through sampling on directed Bayesian models with a few variables. Nessler et al. \cite{nessler2013bayesian} extended this work by incorporating an STDP learning rule to perform Expectation-Maximization on simple Bayesian models where the exact posterior is computationally tractable. Habenschuss et al. \cite{habenschuss2012homeostatic} further extended this framework by incorporating posterior constraints, while Kappel et al. \cite{kappel2014stdp} implemented an approximate form of the Forward-Backward algorithm for Hidden Markov Models using spiking WTA-circuits. Pecevski et al. \cite{pecevski2016learning} considered deeper models but restricted themselves to networks with a small number of hidden states so that exact inference remained computationally tractable. Yu et al. \cite{yu2018winner} relaxed the need for exact inference by considering unstructured mean-field variational inference on undirected Markov-Random fields using spiking WTA circuits, but did not demonstrate learning. 

Guo et al. \cite{guo2017hierarchical} demonstrated unstructured mean-field variational inference and learning on directed probabilistic graphical models (PGMs) with a large number of hidden variables and of arbitrary depth. They thus presented a more scalable approach than the aforementioned methods, at the cost of foregoing \emph{exact} inference. Their choice of an \emph{unstructured} mean-field posterior allowed them to compute the parameters of the variational posterior analytically as a function of the parameters of the generative model. As a result they only had to learn the parameters of the generative model from the data, leading to simpler learning rules. Their scheme requires bidirectional synapses, but they did not address the question of how the feedforward and feedback weights can mirror each other. Furthermore, the limitation of the unstructured mean-field approximation is that it explicitly ignores all correlations between random variables, while conditional probabilities are found to play an important role in experimentally studied task switching and cue integration tasks \cite{cort2013conditional}\cite{schmitz2014components}. It has also been shown that the optimization problem becomes more non-convex under the unstructured mean-field assumption, and conversely, fewer local minima exist for more structured variational posteriors \cite{wainwright2008graphical}. Empirically as well, better parameter estimates are obtained using structured variational posteriors that help to avoid the strong dependence on initial conditions observed in unstructured mean-field inference \cite{hoffman2013stochastic}. This paper builds upon \cite{guo2017hierarchical} by performing \emph{structured} mean-field variational inference on directed PGMs using spiking WTA circuits. Furthermore, instead of \emph{assuming} the existence of bidirectional synapses, we show that our learning rule drives the feedforward and feedback synapses to the same desired value.

Mnih and Gregor \cite{mnih2014neural} presented a framework for structured variational inference and learning. They elaborated on the case when both the generative model and variational posterior are directed PGMs. They learned the parameters of both the PGMs using gradient descent on the variational Free Energy. The gradients take the form of expectations with respect to the variational posterior. They used a separate Monte-Carlo sampler to draw samples from the variational posterior to construct Monte-Carlo estimates for the gradients. Thus they stored the model in memory without explicitly encoding it in a network, and ran a separate Monte-Carlo sampler to compute the required gradients. Algorithmically, our work also fits in this framework as we also perform gradient descent on the variational Free Energy and use the same expressions for the gradients. However, the crucial difference is that we explicitly encode the PGMs in a Spiking Neural Network (SNN), and we compute the required gradients in an online fashion using the spiking dynamics of the SNN. We show that our online learning rules take the form of error-modulated Spike-Timing Depedent Plasticity (STDP).

The well known Helmholtz Machine \cite{dayan1995helmholtz} is the closest model to ours in terms of the choice of graphical models encoding the generative and variational distributions. A small instance of the Helmholtz machine was indeed recently implemented by Sountsov and Miller \cite{sountsov2015spiking} in a network of deterministic spiking neurons. The Helmholtz Machine does however differ in the choice of activation function, which leads to a difference in the interpretation for the synaptic weights as well as significant differences in the dynamics of the network. In particular, the Helmholtz Machine uses an additive neuron model, while the structured mean-field assumption induces a product form on the conditional distributions, and thus we are using a multiplicative neuron model instead. Hence our work is also related to past literature on multiplicative neurons such as pi-sigma networks \cite{shin1991pi}, multiplicative feature integration \cite{wu2018and}, and neural arithmetic logic units \cite{trask2018neural}. We refer the reader to \cite{schmitt2002complexity} for more details on the applications of multiplicative neurons in Machine Learning and Neuroscience. Further, we don't train our model using the wake-sleep algorithm. While the wake-sleep algorithm is found to work well empirically, it does not minimize a single cost function. The other issue with the wake-sleep algorithm is that it requires alternations between a feedforward pass and a backward pass, which undermines its biological plausibility. Instead, we approximate the gradients of the cost function directly in order to derive our learning rule.  	

Our SNN uses stochastic spiking WTA circuits as building blocks. We explore the use of both hard and soft WTA circuits. In hard WTA circuits, the firing of one neuron inhibits all other neurons from firing, whereas in soft WTA circuits each neuron fires independently. We show that using hard WTA circuits leads to the SNN operating in a temporal coding regime and funtioning as a Gibbs sampler for the variational posterior. This sampling based inference then allows us to compute unbiased online Monte-Carlo estimates for the gradients. As noted in \cite{mnih2014neural}, we also find that this approach suffers from the drawback that the Monte-Carlo estimates for the gradients are unbiased but have high variance. However, we do prove asymptotic convergence of our algorithm using stochastic approximation theory. Further, we show that using soft WTA circuits, the SNN operates in a rate-coded regime and performs message passing \cite{yedidia2003understanding} to compute the marginals of the variational posterior. We then derive novel estimates for the gradients using these marginals, and show that while the estimates are biased, they have much lower variance. We thus report a bias-variance tradeoff in these two methods of learning.  This can also be interpreted as a speed-accuracy tradeoff between neural coding strategies wherein the learning is faster but biased with the rate coded soft WTA spiking, but is slower and unbiased with the temporally coded hard WTA spiking. We also note that while message passing allows marginals to be estimated very efficiently, sampling is required to compute higher order moments, and thus these inference strategies are complementary to one another. 

Our scheme differs from previous spiking WTA schemes in that we use a multi-compartment neuron model whereas the previous works used single-compartment models. The neurons in our networks have dendrites that perform logarithmic compression \cite{jones2012logarithmic}\cite{keil2015dendritic}. This is required to implement the sum-product message passing algorithm in our scheme. We thus propose a clear computational role for the dendrites found in a majority of pyramidal neurons in the neocortex.  We note that Rao \cite{rao2005hierarchical} studied inference via belief propagation in networks of spiking neurons with dendrites but only considered shallow networks and did not report any learning. We also note that Hawkins and Ahmad have also recently reported the importance of dendrites in learning with locally competitive circuits in a non-Bayesian setting \cite{hawkins2016neurons}. Our scheme is also compatible with the message passing model for the visual cortex outlined by Mumford and Lee \cite{lee2003hierarchical}. The schemes in \cite{pecevski2011probabilistic}\cite{nessler2013bayesian}\cite{habenschuss2012homeostatic} use sampling while \cite{yu2018winner}\cite{guo2017hierarchical} use message passing. In this work we provide a unified view of these two approaches. In soft WTA circuits, all neurons can fire simultaneously while in hard WTA circuits only one neuron is allowed to fire at a time due to stronger lateral inhibition. Hence switching between hard and soft WTA spiking is simply a matter of changing the amount of lateral inhibition. One plausible way to achieve this is via a globally difussing neuromodulator. To the best of our knowledge, this is the first SNN scheme that offers a unified circuit level view of these two inference approaches.

In section 2 we outline the algorithmic details of structured variational inference and learning. In section 3 we discuss how these computations can be mapped to Spiking Neural Networks with hard WTA circuits. In section 4 we discuss the implementation using soft WTA circuits.

\section{Structured Mean-field Variational Inference and Learning}

Let the generative model that the brain has be $P_{\theta} \left ( x,h \right )$. Here $x$ is the data or the set of sensory inputs, $h$ is the set of hidden states, and $\theta$ is the set of parameters of this model. The problem of inference is to compute the posterior $P_{\theta}(h|x)$. Learning under the maximum likelihood principle corresponds to finding $\theta$ that maximizes the log-likelihood of the data $\sum_{x \in D} \log P_{\theta}(x)$ , where $D$ is the entire dataset or set of sensory inputs. Both these problems require computing the Bayesian surprise $-\log P_{\theta}(x) = -\sum_{h \in H} \log P_{\theta}(x, h)$. This marginalization becomes intractable as the size of the state-space of hidden states denoted by $H$ becomes large. We minimize an upper bound on Bayesian surprise, the variational free-energy, defined as:

\begin{equation}
F\left ( x, \theta, \phi \right ) = -\log P_{\theta}(x) + D_{KL}\left ( Q_{\phi}(h|x) || P_{\theta}(h|x) \right )
\end{equation}

Here $Q_{\phi}(h|x)$ is the variational posterior that we use to approximate the true posterior, parametrised by $\phi$. In the rest of the paper, we denote the free-energy $	F\left ( x, \theta, \phi \right )$ simply as $	F\left ( x \right )$ for brevity of notation. The gradients of the free-energy w.r.t to $\theta$ and $\phi$ are: 

\begin{align}
\nabla_{\theta} F(x) &= \mathbb{E}_Q \left [-\nabla_{\theta} \log P_{\theta}(x,h) \right ] \\
\nabla_{\phi} F(x) &= \mathbb{E}_Q \left [ \left ( \log P_{\theta}(x,h) - \log Q_{\phi}(h|x) \right ) \cdot \left ( -\nabla_{\phi} \log Q_{\phi}(h|x) \right ) \right ]
\end{align}

Note that the expectations in (2) and (3) are with respect to the variational posterior $Q_{\phi}(h|x)$. We refer the reader to the appendix of \cite{mnih2014neural} for a derivation of (2) and (3). In this paper the generative models that we consider are tree-structured directed PGMs (see Fig.1a). This is because trees capture hierarchical structure and other Bayesian models can be converted to a tree by clustering variables together \cite{yedidia2003understanding} and this also motivated the same choice in \cite{guo2017hierarchical}. We index the nodes in the tree from $1$ to $N$, with $z_c$ denoting the random variable corresponding to the node with index $c$. The data corresponds to the leaves of the tree as $x = \{z_1,z_2,...,z_M\}$. The remaining nodes comprise the hidden states as $h = \{z_{M+1},z_{M+2},...,z_N\}$. Let $pa(c)$ denote the index of the parent node of node c in the tree. Further, let each $z_c$ take (utmost) $K$ values. Then the generative model factors out top-bottom as:

\begin{equation}
P_{\theta}(x,h) = p_{\theta}(z_N)\prod_{c=1}^{N-1}\prod_{j=1}^{K}\prod_{i=1}^{K}\left ( p_{\theta}(z_c=i|z_{pa(c)}=j) \right )^{\delta(z_c=i) \cdot \delta(z_{pa(c)}=j)}
\end{equation}

Here $\delta(z_c = i)$ denotes the delta function which is equal to $1$ if $z_c=i$ and $0$ otherwise. Further, let $ch(c)$ denote the set of children nodes of node c in the tree corresponding to the generative model. Thus we have $ch(c) = ch(c)^0, ch(c)^1, ..., ch(c)^R$ for a node $c$ with $R$ child nodes. The structured variational posterior that we choose to work with in this paper corresponds to simply inverting the direction of the arrows in the generative model (see Fig.1b). This allows us to retain an intuitive and reasonable set of dependencies between random variables in the variational posterior and is similar to the choice made in \cite{mnih2014neural}. For this choice of variational posterior, the value of node $c$ is conditionally dependent on the set of values of its child nodes in the generative model. We further assume a fully factored form for the conditional probabilities in our variational posterior. This yields a structured mean-field variational posterior, that factors in a feed-forward manner as:

\begin{equation}
Q_{\phi}(h|x) = \prod_{c=M+1}^{N}\prod_{r=1}^{R}\prod_{j=1}^{K}\prod_{i=1}^{K}\left ( q_{\phi}(z_c=i|z_{ch(c)^r}=j) \right )^{\delta(z_c=i) \cdot \delta(z_{ch(c)^r}=j)}
\end{equation}

Note that $ch(c)$ used in (5) is defined with respect to the tree for the generative model, and not with respect to the (non-tree) graph for the variational posterior. Thus it is important to note that for consistency of notation, all subsequent use of $pa(c)$ and $ch(c)$ will always be with respect to the tree corresponding to the generative model (see Fig.1a). Note also that while the LHS of (5) explicitly contains the data variables $x$, the RHS does not. This is because they are subsumed in the set of child nodes for the other variables, and don't have child nodes themselves. This is also why the index $c$ for the nodes starts from $M+1$ in (5), noting again that the data nodes have indices from $1$ to $M$.   

\begin{figure}[t]
	\centering
	\includegraphics[width=\linewidth]{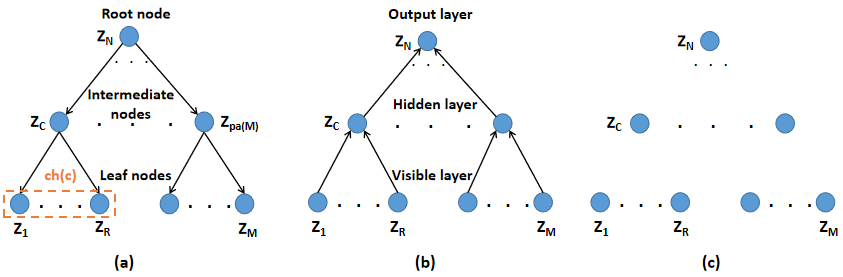}
	\caption{(a) Graphical representation of the generative model. The arrows depict  dependencies between variables. The parent node for $z_M$ has been marked as $z_{pa(M)}$ and the child nodes of $z_c$ have been encircled as $ch(c)$. (b) The structured mean-field variational posterior. Note that the arrows are flipped with respect to Fig.1a (c) The unstructured mean-field variational posterior for reference. The lack of arrows indicates that all nodes are independent of each other.}
\end{figure}

The conditional distributions in (5) were chosen to be fully factored because for the non-factored case there is a multiplicative blowup in the number of terms (and hence synapses in the SNN). To see this, consider a node $c$ with two child nodes $a$ and $b$. If $K=10$ for both $a$ and $b$, then we will require $K \times K = 100$ terms for each value that $c$ can take. On the other hand, by assuming a factored form for the conditional distribution, we only require $K + K = 20$ terms, and thus there is an additive increase in the number of terms (and hence synapses) in this case. Furthermore, in the next section we show that the factored form for the conditional distribution also greatly simplifies the Markov blanket of the random variables in the graph, and thus allows for a more compact network with fewer synapses. We also note that the structured mean-field still retains conditional dependencies between variables, as can be seen from the expression in (5), and is not the same as the unstructured (i.e fully factorized) mean-field model (see Fig.1c). We will drop $\theta$ and $\phi$ from expressions for the distributions in subsequent sections for brevity. Note that as we will encode these probability distributions using SNNs, and hence $\theta$ and $\phi$ will simply correspond to the set of synaptic weights of these networks. Spiking nonlinearity of neurons is modeled through a WTA mechanism in the next section, in which we do not introduce extra parameters. 

\newpage

\section{Implementation by spiking neurons with hard WTA circuits}

\subsection{Gibbs sampling via hard WTA spiking}

For every random variable $z_c$ that can take $K$ values in the variational posterior (5), there is a corresponding winner-take-all (WTA) circuit $G_c$ comprised of $K$ neurons in our SNN (see Fig. 2a). These $K$ neurons are indexed as $z_c^1,z_c^2,...,z_c^K$ and have corresponding membrane potentials $u_c^1(t),u_c^2(t),...,u_c^K(t)$ that vary with time $t$. The firing rate of neuron $z_c^i$ at time $t$, denoted by $\rho_c^i(t)$ is given as:

\begin{equation}
\rho_{c}^i(t) = \frac{\exp(u_{c}^i(t)))}{\sum_{k=1}^K \exp(u_{c}^k(t))}
\end{equation}

\begin{figure}[t]
	\centering
	\includegraphics[width=0.5\linewidth]{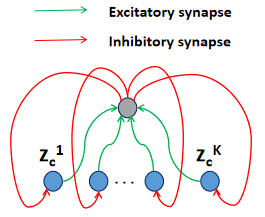}
	\caption{Diagram of WTA circuit $G_c$ corresponding to random variable $z_c$. The $K$ neurons (in blue) that each encode one of the $K$ possible values of $z_c$ are connected via excitatory synapses to a common neuron (in grey), which upon firing of any of the $K$ neurons inhibits all of them firing for a while by injecting negative current via inhibitory synapses. cf. Fig. 1 in \cite{guo2017hierarchical}.}
\end{figure}

In this section we are working with hard WTA circuits, which have stronger lateral inhibition than the softer WTA circuits that we discuss in the next section. Let neuron $z_c^i$ fire at time $t$, indicating that the random variable $z_c$ has switched to state $i$ at time $t$. We denote the state of the random variable $z_c^i$ at time $t$ by $z_c^i(t)$. In this hard WTA circuit, all the $K$ neurons are inhibited from firing for a refractory period $\tau$ after this firing event, meaning that $z_c(t) = i$ between $t$ and $t+\tau$. After this refractory period, $z_c$ can again switch states, as governed by (6). Let $S_c^i(t) = \sum_{t_f \in F_c^i(t)}\delta_D(t-t_f)$ denote the spike-train of neuron $z_c^i$, with $F_c^i(t)$ being the set of time instants when $z_c^i$ fired until time $t$. Further note that $\delta_D(t-t_f)$ is a Dirac-delta function centered at $t_f$, with the subscript $D$ being used to distinguish it from the Kronecker-delta functions used in (4) and (5). Firing of $z_c^i$ leads to the injection of current $I_c^i(t)$ into postsynaptic neurons, weighted by the corresponding synaptic weights. $I_c^i(t)$ is modeled as a filtered version of $S_c^i(t)$. The ideal rectangular spike-response kernel is softened (Fig.3) to a more biologically realistic double-exponential window $\kappa(t)$:

\begin{equation}
\kappa(t) = \kappa_0 \cdot \left ( \exp\left ( \frac{-t}{\tau_f} \right ) - \exp\left ( \frac{-t}{\tau_s} \right ) \right )
\end{equation}

Here $\kappa_0$ is a constant to scale the maximum value of $\kappa(t)$ to $1$ and $\tau_f$, $\tau_s$ denote the two timescales that parametrize the kernel. $I_c^i(t)$ is then given as:

\begin{equation}
I_c^i(t) = \int_{0}^{t}\kappa(s) \cdot S_c^i(t-s)ds
\end{equation}

\begin{figure}[t]
	\centering
	\includegraphics[width=0.6\linewidth]{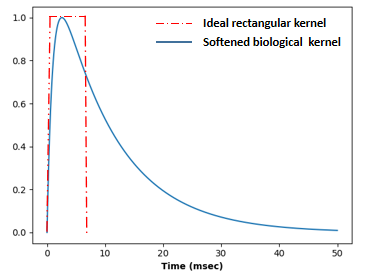}
	\caption{The spike-response kernels. The ideal rectangular kernel has zero rise time and fall time. The more biologically realistic kernel is modeled as a double-exponential and has finite rise time and fall time.}
\end{figure}

If $z_r$ feeds into $z_c$ (thus $r \in ch(c)$), then let $\phi_r^{ij}$ denote the weight of the synapse connecting $z_r^j$ to $z_c^i$. Note that we are working with bidirectional synapses in this section, meaning that we assume there is a synapse from $z_c^i$ to $z_r^j$ of the same magnitude as the one connecting $z_r^j$ to $z_c^i$. We will discuss how we can achieve this mirroring of synaptic weights in a section 3.3. These synaptic weights are used to encode terms from (4) and (5) as:

\begin{align}
\theta_c^{ij} &= p(z_c=i|z_{pa(c)}=j) \\
\phi_r^{ij} &= q(z_c=i|z_r=j)
\end{align}

\begin{figure}[t]
	\centering
	\includegraphics[width=0.6\linewidth]{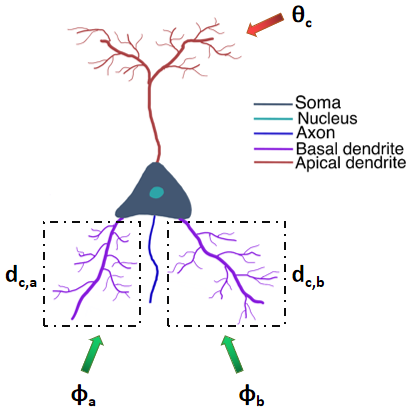}
	\caption{Diagram for one of the neurons in $G_c$. Here $\phi_a$ and $\phi_b$ respectively denote the set of feedforward synapses incident upon this neuron from $G_a$ and $G_b$. Note that $\phi_a$ and $\phi_b$ are respectively incident on two separate basal dendritic branches $d_{c,a}$ and $d_{c,b}$. The set of feedback synapses from the generative model, denoted by $\theta_c$, are incident on the apical dendrite.}
\end{figure}

We posit that the weights for the variational posterior given by (10) (along with their mirrored copies along backward connections) are incident on basal dendrites, while weights for the generative model given by (9) are incident on apical dendrites (see Fig.4), which is consistent with previously proposed models for cortical microcircuitry \cite{hawkins2016neurons}\cite{sacramento2018dendritic}. However in this paper, we only model the current dynamics due to the basal dendrites. This is a reasonable first approximation as the apical dendrites are known to inject much smaller amounts of currents compared to basal dendrites. We can then rewrite (4) and (5) in terms of the synaptic weights as: 

\begin{align}
P_{\theta}\left ( x,h \right ) &= p(z_N)\prod_{c=1}^{N-1}\prod_{j=1}^{K}\prod_{i=1}^{K}\left ( \frac{1}{Z_c^{'j}}\left ( \theta_{c}^{ij} \right )^{\delta(z_c=i)} \right)^{\delta(z_{pa(c)}=j)} \\
Q_{\phi}\left ( h|x \right ) &= \prod_{c=M+1}^{N}\prod_{r=1}^{R}\prod_{j=1}^{K}\prod_{i=1}^{K}\left ( \frac{1}{Z_r^j}\left ( \phi_{r}^{ij} \right )^{\delta(z_c=i)} \right)^{\delta(z_r=j)} 
\end{align}

Here $Z_c^{'j} = \sum_{i=1}^K\theta_c^{ij}$ and $Z_r^j = \sum_{i=1}^K\phi_r^{ij}$ are normalization constants. We assume that the weights are initialised such that all these normalization constants are equal to $1$, thus ensuring that the conditional distributions defined by (9) and (10) are normalised. We subsequently ensure that the values of these normalization constants do not change over the course of learning. Hence we omit these constants from the expressions while discussing inference and only consider them while deriving the learning rules. Equations (11) and (12) then simplify to:  

\begin{align}
P_{\theta}\left ( x,h \right ) &= p(z_N)\prod_{c=1}^{N-1}\prod_{j=1}^{K}\prod_{i=1}^{K} \left ( \theta_c^{ij} \right ) ^{\delta(z_c=i) \cdot \delta(z_{pa(c)}=j)} \\
Q_{\phi}\left ( h|x \right ) &= \prod_{c=M+1}^{N}\prod_{r=1}^{R}\prod_{j=1}^{K}\prod_{i=1}^{K}\left ( \phi_{r}^{ij} \right)^{\delta(z_c=i) \cdot \delta(z_r=j)} 
\end{align}

We now proceed to describe the dynamics of the neuron membrane potentials. Consider the random variable $c$ and let it have $R$ children i.e $|ch(c)| = R$, and has one parent node due to the tree structure of the network. Then each neuron $z_c^i$ ($i \in [1,K]$) has $R+1$ basal dendrites. The first $r$ basal dendrites have incident synaptic currents exclusively from the respective $r$'th child node. The $(R+1)$'th dendrite has currents incident from the parent node with weights that mirror the feedforward weights from this node to the parent node, and hence in what follows, $z_{R+1} \equiv z_{pa(c)}$ and $\phi_{R+1}^{ij} \equiv \phi_{c}^{ji} = q(z_{pa(c)} = j | z_c = i)$. There is thus a clear topological segregation of inputs in the dendritic tree, with one dendrite collecting current from the neurons encoding one random variable. Each of these dendrites is assumed to be coupled to the soma by a unit resistance. Further, each dendrite outputs the logarithm of the current incident on it. Let $d_{c,r}^i(t)$ denote the current from the $r$'th basal dendrite, which is the current collected from neurons encoding $z_r$. We then have:

\begin{align}
u_c^i(t) &= \sum_{r=1}^{R+1}d_{c,r}^i(t) \\
d_{c,r}^i(t) &= \log \left ( \sum_{j=1}^{K}I_r^j(t) \cdot \phi_r^{ij} \right )
\end{align}

\begin{figure}[t]
	\centering
	\includegraphics[width=\linewidth]{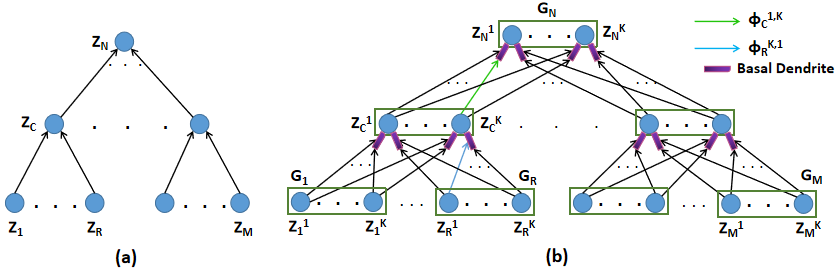}
	\caption{(a) The variational posterior defined by (5). (b) The corresponding SNN: For each variable $z_i$ in (a) that can take $K$ values, there is a WTA circuit $G_i$ with $K$ neurons in (b). Two synapses have been colored and named to illustrate how synapses are connected as per (10). Note that for every feedforward synapse depicted here, there is also a symmetric feedback connection (not depicted in the figure) that is also incident on a basal dendrite. cf. Fig. 2 in \cite{guo2017hierarchical}.}
\end{figure}

Having described the spiking and current dynamics of our SNN, we proceed to show that this network performs Gibbs sampling on the variational posterior. We start by noting that the spike-response kernel defined in (7) has a peak value of $1$, and for a time-interval around this peak, its value is close to $1$. Hence, $I_c^i(t)$ has a value close to $1$ after neuron $z_c^i$ fires (and is $0$ otherwise). We also note that only one of the neurons in a WTA circuit is active at a time due to the refractory period. This ensures that the random variable $z_c$ takes a unique value at any instant in time. We can then also see that $I_c^i(t)$ approximates $\delta(z_c(t)=i)$, with $z_c(t)$ being the state of the random variable $z_c$ at time $t$. Combining this observation with equations (9), (10), (11), (12) and substituting in (6), we get:

\begin{equation}
\begin{aligned}
\rho_{c}^i(t) & \propto \exp(u_c^i(t)) = \exp(\sum_{r=1}^{R+1}d_{c,r}^i(t)) = \exp\left ( \sum_{r=1}^{R+1}\log \left ( \sum_{j=1}^{K}I_r^j(t) \cdot \phi_r^{ij} \right ) \right ) \\
& \approx \exp\left ( \sum_{r=1}^{R+1}\log \left ( \sum_{j=1}^{K}\delta(z_r(t) = j) \cdot \phi_r^{ij} \right ) \right ) = \prod_{r=1}^{R+1}\left ( \sum_{j=1}^{K}\delta(z_r(t) = j) \cdot \phi_r^{ij} \right ) \\
& = \left ( \sum_{j=1}^{K}\delta(z_r(t) = j) \cdot q(z_{pa(c)}=j|z_c=i)\right ) \cdot \prod_{r=1}^{R}\left ( \sum_{j=1}^{K}\delta(z_r(t) = j) \cdot q(z_c=i|z_r=j)\right ) \\
& = \prod_{j=1}^{K}q(z_{pa(c)}=j | z_c=i)^{\delta(z_{pa(c)}(t) = j)} \cdot \prod_{r=1}^{R}\prod_{j=1}^{K}q(z_c=i|z_r=j)^{\delta(z_r(t) = j)}
\end{aligned}
\end{equation}

Note that the last equality in line 4 of (17) is due to the fact that exactly one of the $\delta(z_r(t)=j)$, $r \in R+1$ is equal to $1$ at any time while the others are $0$, due to the aforementioned fact that each of the random variables takes a unique value at time $t$. A Gibbs sampler works by changing the state of one variable in the network at a time. The new state for the chosen variable is sampled from the conditional distribution obtained by keeping all the other variables in the network fixed at their current values. It is fairly straightforward to see that (17) is this desired conditional distribution. This is because the RHS is the product of all factors in the distribution that are a function of $z_c$. This can also be worked out by simplifying the Bayes rule expression for the condtional distribution in (18).

\begin{equation}
\begin{aligned}
Q_{\phi}(z_c|(h \setminus  z_c), x) &= \frac{Q_{\phi}(z_c, (h \setminus  z_c)| x)}{Q_{\phi}((h \setminus  z_c)| x)} \\
=\frac{Q_{\phi}(h| x)}{\sum_{z_c}Q_{\phi}(h| x)} &= \frac{\prod_{a=M+1}^{N}\prod_{r=1}^{R}q_{\phi}(z_a|z_{ch(a)^r})}{\sum_{z_c}\prod_{a=M+1}^{N}\prod_{r=1}^{R}q_{\phi}(z_a|z_{ch(a)^r})}
\end{aligned}
\end{equation}

A very important point to note here is that for a general graphical model, the markov blanket of a random variable comprises it's children nodes, parents nodes and the other parents of its children nodes. However we find that (17) does not contain terms with the other parents of the child nodes, which happens precisely as a consequence of our structured mean-field assumption. This considerably simplified Markov blanket allows for fewer synapses in the network as well as simpler dynamics and weight updates. Also note that the soft-max nature of the WTA circuit ensures that the $\rho_c^i(t)$ sum up to $1$ to form a probability distribution. As the SNN operates in continuous time, the probability of any two neurons in the network firing at the exact same time is $0$. This implies that variables change their states one at a time. Further, the waiting time for the next spike from the WTA circuit of any variable $z_c$ that is not in its refractory period, is exponentially distributed with rate parameter $\lambda = \sum_{i=1}^{K}\rho_c^i(t) = 1$ and hence the next variable to switch its state is being chosen uniformly at random. Thus we conclude that the network is a Gibbs sampler for the variational posterior, operating in continuous time. We note that this result also holds with the inclusion of refractory periods as each WTA circuit has the same refractory period.	

\subsection{Learning via sampling}

The gradients of the free-energy given by (2) and (3) take the form of expectations of certain functions over the variational posterior. A closed-form expression for these gradients is not known. However, we can estimate these gradients via Monte-Carlo sampling. In the previous section we showed that our SNN is a Gibbs sampler for the variational posterior and generates samples, one at a time. To allow for real-time weight updates, akin to experimentally observed STDP, we use only one sample in our Monte-Carlo estimate, doing away with the need for accumulating samples from the network and then making subsequent weight updates. Using just one sample gives us a noisy estimator of the true gradient. But it is important to note that the expected value of this estimator is indeed the true gradient, as our network is a Gibbs sampler for the variational posterior. Given this unbiased estimator for the true gradient, we proceed to formulate our learning rule as an instance of the Robbins-Monro algorithm \cite{borkar2013stochastic}. Let $\Delta \phi_r^{ij}$ denote the estimated gradient of the free-energy with respect to synaptic weight $\phi_r^{ij}$ using the current sample from the SNN. Also, let $\Delta \theta_c^{ij}$ be the corresponding change in $\theta_c^{ij}$. We perform stochastic gradient-ascent on the negative free-energy (or equivalently gradient descent on the free-energy) by updating weights as:

\begin{align}
\phi_r^{ij} &\leftarrow \phi_r^{ij} + \xi_r^{ij}(t) \cdot \Delta \phi_r^{ij} \\
\theta_c^{ij} &\leftarrow \theta_c^{ij} + \xi_c^{'ij}(t) \cdot \Delta \theta_c^{ij}
\end{align}

Correctness of this stochastic scheme is guaranteed \cite{borkar2013stochastic} for learning rates that satisfy:

\begin{equation}
\lim_{t \rightarrow \infty}  \xi(t) = 0, \sum_{t = 1}^{\infty}  \xi(t) = \infty, \sum_{t = 1}^{\infty} \left ( \xi(t) \right )^2 = \infty 
\end{equation}

Note that $\xi(t)$ in (21) refers to all the $\xi_r^{ij}(t)$ and $\xi_c^{'ij}(t)$ specified in (19) and (20). A specific choice of learning rates that satisfy (21) is $\xi_r^{ij}(t) = \frac{1}{N_r^j(t)}$ and $\xi_c^{'ij}(t) = \frac{1}{N_c^i(t)}$, where $N_r^j(t)$ and $N_c^i(t)$ are respectively the number of spikes of the neuron $z_r^j$ and $z_c^i$ until time $t$. As noted before, while we omitted the normalization constants while discussing inference, they need to be considered while deriving the learning rules. Thus we differentiate (12) to get:

\begin{align}
\nabla_{\theta_c^{ij}}\log P_{\theta}(x,h) &= \frac{1}{\theta_c^{ij}}\delta(z_c=i) \cdot \delta(z_{pa(c)}=j) - \frac{1}{Z_c^{'j}}\delta(z_{pa(c)}=j) \\
\nabla_{\phi_r^{ij}}\log Q_{\phi}(h|x) &= \frac{1}{\phi_r^{ij}}\delta(z_c=i) \cdot \delta(z_r=j) - \frac{1}{Z_r^j}\delta(z_r=j)
\end{align}

Define $e(x,h) = \log P_{\theta}(x,h) - \log Q_{\phi}(h|x)$. The gradients with respect to the negative of the Free Energy are then given as:

\begin{align}
-\nabla_{\theta_c^{ij}}F(x) &= \mathbb{E}_{Q}\left [ \frac{1}{\theta_c^{ij}}\delta(z_c=i) \cdot \delta(z_{pa(c)}=j) - \frac{1}{Z_{c}^{'j}}\delta(z_{pa(c)}=j) \right ] \\
-\nabla_{\phi_r^{ij}}F(x) &= \mathbb{E}_{Q}\left [e(x,h) \cdot  \left ( \frac{1}{\phi_r^{ij}}\delta(z_c=i) \cdot \delta(z_r=j) - \frac{1}{Z_r^j}\delta(z_r=j) \right ) \right ]
\end{align}

As mentioned before, we use Monte-Carlo estimates of (24) and (25) in our learning rules, and only a single sample is used to allow for real-time updates. This then yields the following weight update formulae:  

\begin{align}
\Delta \theta_c^{ij} &= \frac{1}{\theta_c^{ij}}\delta(z_c=i) \cdot \delta(z_{pa(c)}=j) - \frac{1}{Z_{c}^{'j}}\delta(z_{pa(c)}=j) \\
\Delta\phi_r^{ij} &= e(x,h) \cdot  \left ( \frac{1}{\phi_r^{ij}}\delta(z_c=i) \cdot \delta(z_r=j) - \frac{1}{Z_r^j}\delta(z_r=j) \right )
\end{align}

We had derived equation (13) and (14) from (11) and (12) by assuming that all the normalization constants $Z_r^j$ and $Z_{c}^{'j}$ were equal to $1$. We had assumed that the weights were initialised such that this was true. We now proceed to rescale (26) and (27) so that these constants remain equal to $1$ after the weight updates. We multiply (26) and (27) respectively by $\theta_c^{ij}$ and $\phi_r^{ij}$ and use the fact that the normalization constants are initially equal to $1$ to get new update rules as:

\begin{align}
\Delta \theta_c^{ij} &= \delta(z_c=i) \cdot \delta(z_{pa(c)}=j) - \theta_c^{ij} \cdot \delta(z_{pa(c)}=j) \\
\Delta\phi_r^{ij} &= e(x,h) \cdot  \left ( \delta(z_c=i) \cdot \delta(z_r=j) - \phi_r^{ij} \cdot \delta(z_r=j) \right )
\end{align}	

Note that these rescaled gradients preserve the sign of the original gradient as we have multiplied by positive numbers. Furthermore, the rescaling does not change the conditions for stationary points (obtained by setting all these weight updates to $0$) and hence they will also converge to the same set of optima as (26) and (27). We now proceed to show that (28) and (29) indeed preserve the values of the normalization constants.

\begin{align}
&\begin{aligned}
\Delta Z_c^{'j} &= \sum_{i=1}^{K}\Delta \theta_c^{ij} = \sum_{i=1}^{K}\left ( \delta(z_c=i) \cdot \delta(z_{pa(c)}=j) - \theta_c^{ij} \cdot \delta(z_{pa(c)}=j) \right ) \\
&= \delta(z_{pa(c)}=j) \cdot \left ( \sum_{i=1}^{K} \delta(z_c=i) - \sum_{i=1}^{K} \theta_c^{ij} \right ) = \delta(z_{pa(c)}=j) \cdot (1 - 1) = 0
\end{aligned} \\
&\begin{aligned}
\Delta Z_r^{j} &= \sum_{i=1}^{K}\Delta \phi_r^{ij} = \sum_{i=1}^{K} e(x,h) \cdot  \left ( \delta(z_c=i) \cdot \delta(z_r=j) - \phi_r^{ij} \cdot \delta(z_r=j) \right ) \\
&= e(x,h) \cdot \delta(z_r=j)  \cdot \left ( \sum_{i=1}^{K}  \delta(z_c=i) - \sum_{i=1}^{K}  \phi_r^{ij} \right ) = e(x,h) \cdot \delta(z_r=j) \cdot \left ( 1 - 1 \right )  = 0
\end{aligned}
\end{align}

The last step in (30) and (31) used the facts that $\sum_{i=1}^{K} \theta_c^{ij} = Z_c^j = 1$ and $\sum_{i=1}^{K}  \phi_r^{ij} = Z_r^j = 1$. They also used the fact that $\sum_{i=1}^{K}  \delta(z_c=i) = 1$, which is true as $z_c$ takes exactly one value at a time. The terms $\delta(z_c=i) \cdot \delta(z_r=j)$ and $\delta(z_c=i) \cdot \delta(z_{pa(c)}=j)$ show that these are Hebbian learning rules, requiring the detection of concurrent firing in the presynaptic and postsynaptic neuron. To implement this in real-time in our SNN, we set $\delta(z_c=i) \cdot \delta(z_r=j) = 1$ if neuron $z_r^j$ fired and then postsynaptic neuron $z_c^i$ also fired within time $\sigma$ of this presynaptic firing event. The same logic is used to to estimate $\delta(z_c=i) \cdot \delta(z_{pa(c)}=j)$. This yields a rectangular STDP window, which we approximate (as for the spike-response kernel, Fig.2) using a more biologically realistic double-exponential window $W(t)$:

\begin{equation}
W(t) = w_0 \cdot \left ( \exp\left ( \frac{-1}{\tau_a} \right ) - \exp\left ( \frac{-1}{\tau_b} \right ) \right )
\end{equation}

As with $\kappa_0$ in (7), $w_0$ is a constant to scale the maximum value of $W(t)$ to $1$ and $\tau_a$, $\tau_b$ denote the two timescales that parametrize this kernel. This then yields the STDP learning rules:

\begin{align}
\Delta \theta_c^{ij} &= S_c^i(t) \cdot \int_{0}^{t}S_{pa(c)}^j(t-s) \cdot W(s)ds - \theta_c^{ij} \cdot S_{pa(c)}^j(t) \\
\Delta\phi_r^{ij} &= e(x,h) \cdot  \left ( S_c^i(t) \cdot \int_{0}^{t}S_r^j(t-s) \cdot W(s)ds - \phi_r^{ij} \cdot S_{r}^j(t) \right )
\end{align}	

We note that this idea of rescaling the gradients was introduced in \cite{nessler2013bayesian} and was also subsequently used in \cite{guo2017hierarchical}. The use of Lagrange multipliers to ensure normalization of the weights as an alternative to rescaling the gradients was also explored in \cite{nessler2013bayesian}, and the same approach can also be used in our framework. However, these previous approaches encoded the logarithm of the probabilities directly in the synaptic weights as they did not have dendrites that perform logarithmic compression. As a result, they had to make a Taylor series approximation in their rescaled gradients, whereas here the rescaling is exact. Further, as they were encoding the logarithm of the probabilities, the weights in their network could take negative values. They attempted to address this issue by adding a positive offset to all the synaptic weights in the network. However we note that for small enough probabilities, the logarithm of this probability will tend to infinity and the corresponding synaptic weight will be negative, despite the offset. Hence another advantage of our formulation is that our synaptic weights are always positive and bounded. We note however, that the most important advantage of these dendrites will become clear in the next section when we discuss their role in implementing message passing with soft WTA circuits. 

\subsection{Learning bidirectional weights using weight decay}

In Section 3.2, we learned the weights of the synapses encoding the variational posterior while assuming that these synapses are bidirectional. A bidirectional synapse means that we have one feedforward and one feedback synapse with the same weight. If the two synapses were initialized with the same weight, then we can use identical (weight-dependent) STDP updates to ensure that they remain equal, functioning effectively as a bidirectional synapse. In this section, we wish to address the case when they are not initialized to the same value by modifying our STDP learning rule so that the new learning rule drives the feedforward and feedback synaptic weights to rapidly become equal over time. The idea of identical weight changes, with no weight decay, was explored in a paper on mirrored STDP \cite{burbank2015mirrored} to learn a simple two-layer autoencoder. The idea of using weight-decay was first proposed by Kolen and Pollack in 1994 \cite{kolen1994backpropagation} and has more recently resurfaced in the work of Lillicrap et. al \cite{akrout2019deep}, where they attempt to build a “weight mirror” for a biologically plausible Deep-Learning framework. 

Let $w_f$ and $w_b$ respectively be the feedforward and feedforward synapses encoding the term $\phi_r^{ij}$ the variational posterior, and further define $d = w_f - w_b$. We thus want $d=0$ for weight alignment and further that $w_f = w_b = q(z_c=i|z_r=j)$ for convergence to the correct value. Our modified learning rule is obtained by adding a weight-decay term to (29):

\begin{equation}
\begin{gathered}
\Delta\phi_f = \frac{e(x,h)}{N_r}\left ( \delta(z_c = i) \cdot \delta(z_r = j) - \phi_f \cdot \delta(z_r = j) \right ) - \lambda \cdot \phi_f \cdot \delta(z_r = j) \\
\Delta\phi_b = \frac{e(x,h)}{N_r}\left ( \delta(z_c = i) \cdot \delta(z_r = j) - \phi_b \cdot \delta(z_r = j) \right) - \lambda \cdot \phi_b \cdot \delta(z_r = j)
\end{gathered}
\end{equation} 

Here $N_r$ is the number of times $z_r^i$ has fired so far. We assume that the global error term $e(x,h)$ is normalized between $-1$ and $1$. Further, we impose $\lambda > \frac{1}{N_r}$, which implies that $\frac{e(x,h)}{N_r} + \lambda > 0$. Consider now an instant when $z_r$ fires and thus $\delta(z_r = j) = 1$. We then have:

\begin{equation}
\begin{aligned}
\Delta d &= - (\phi_f - \phi_b) \cdot (\frac{e(x,h)}{N_r} + \lambda) \\
d + \Delta d &= d \cdot (1 - (\frac{e(x,h)}{N_r} + \lambda)) \\
&\Rightarrow |d + \Delta d| < |d|
\end{aligned}
\end{equation}

We thus see that each time $z_r$ fires, $w_f$ and $w_b$ are updated and the difference between their values, $d$, goes down each time. Note that while the rate of weight alignment increases with the value of $\lambda$, the distortion in our gradient estimates also increases with $\lambda$. Hence, once $d$ is small enough, the weight decay term can be set to zero to ensure that $w_f$ and $w_b$ indeed converge to the correct values.

\section{Implementation by spking neurons with soft WTA circuits}

\subsection{Message Passing via soft WTA spiking}

In this section we show that by using soft WTA circuits instead of the hard WTA circuits used in the previous section, the dynamics of the network can be shown to be implementing the feedforward pass of belief propagation over the variational posterior. Thus the marginals of the root node are computed exactly, while the marginals of the other random variables are computed approximately under this scheme. An important point to note here is that in this section, we use directed feedforward synapses instead of the bidirectional synapses used in the previous section. Belief propagation is a specific instance of a message passing algorithm that efficiently computes the marginals of graphical models \cite{yedidia2003understanding}. In many cases, it is these marginals that are of interest. We proceed to prove the proposed correspondence and then show how the gradients required for learning can also be estimated using these marginals, thus providing an alternative to the Monte-Carlo sampling approach described in the previous section with hard WTA spiking. We now proceed to describe the feedforward pass of the message passing scheme. Let $q_f(z_c)$ denote the feedforward estimate for the marginal distribution of the variable $z_c$. Then $q_f(z_c)$ is obtained by passing messages forward on the subgraph obtained by removing all the child nodes of $z_c$, so that it is now the root node. Thus we see that $q_f$ and $q$ will agree only for the root node of the original graphical model. Note that for certain applications, such as the unsupervised MNIST classification task as setup in \cite{guo2017hierarchical}, computing just this final marginal correctly suffices. We then see that $q_f(z_c)$ is recursively given as:

\begin{equation}
q_f(z_c = i,z_{ch(c)^1} = i_1,...,z_{ch(c)^R} = i_R) = \prod_{r=1}^{R}\left ( q(z_c = i|z_{ch(c)^r} = i_r)q_f(z_{ch(c)^r} = i_r) \right )
\end{equation}

Here $q_f(z_{ch(c)^k} = i_k)$ is the marginal probability for random variable $z_{ch(c)^k}$ taking value $i_k$ and $R = \left | ch(c) \right|$, as computed using only feedforward messages. We then wish to compute the marginal $q_f(z_c = i)$, which is obtained by summing up (36) over all possible values of the children nodes. Belief propagation is also called the sum-product algorithm as it involves expressing the desired marginal as a product of messages, with each message being computed as a weighted sum of previously computed marginals, and can thus be seen as an efficient dynamic programming approach. In this instance we have:

\begin{equation}
\begin{aligned}
q_f(z_c = i) &= \sum_{i_1,...,i_R}\prod_{r=1}^{R}\left ( q(z_c = i|z_{ch(c)^r} = i_r)q_f(z_{ch(c)^r} = i_r) \right ) \\
&= \prod_{r=1}^{R}\left (\sum_{i_r} \left ( q(z_c = i|z_{ch(c)^r} = i_r)q_f(z_{ch(c)^r} = i_r) \right )  \right ) 
\end{aligned}
\end{equation}

Computing the $q(z_c = i)$ instead of $q_f(z_c = i)$ is a much more challenging task with neural circuitry. This can be done by implementing the full-blown version of Belief Propagation (BP) that also has feedback messages. The problem is that while neurons can only communicate using spikes, BP requires a different message to be sent on each outgoing edge. One way to do this is to use extra neurons to encode each of the messages separately. Another interesting way to address this issue is to use an algorithm called Tree Based Reparametrization \cite{wainwright2003tree}, which reparemetrizes the distribution in terms of the second and first order marginals and estimates these marginals directly without the use of directional messages. This has also recently been implemented at the level of neural population codes by Raju and Pitkow \cite{raju2016inference}. Another approach is to use a different, more easily implementable message passing scheme to obtain \emph{approximate} marginals, instead of the exact marginals obtained using Belief Propagation. While it is true that the feedforward message passing scheme described above trivially provides approximate marginals, it should be possible to incorporate feedback messages for improved estimates. Parr et al have recently proposed one such message passing scheme called Marginal Message Passing \cite{parr2019neuronal}. In the derivation for the gradients that follows, we remain agnostic to the method used to estimate the marginals. 

We have previously been working with hard WTA circuits, wherein only one neuron is allowed to fire at a time. Now we instead consider soft WTA circuits, wherein each neuron in the circuit fires as an independent Poisson process. The firing of one neuron does not inhibit the other neurons in the circuit from firing at the same time, and there is also no refractory period. In this soft WTA spiking regime, we also use a different scaling constant $\kappa_0$ for the EPSP kernel given by (7). Instead of choosing $\kappa_0$ so that the peak value of the kernel is $1$, we now choose this constant so that $\int_{0}^{\infty}\kappa(t)dt = 1$. We note that this condition for the kernel was also stated in \cite{guo2017hierarchical}, but without the subsequent theoretical analysis that we detail here. It can be easily verified that the corresponding choice for this constant is $\kappa_0 = \frac{1}{\tau_f-\tau_s}$.  Note that for this choice of the scaling constant, the peak value of the kernel is much larger than the previous peak value of $1$, which is consistent with the lowered inhibition in this soft spiking regime. Thus we model the effect of lowering the lateral inhibition as threefold: all neurons in the WTA circuit are now allowed to fire concurrently as the firing of one neuron does not inhibit the others, there is no refractory period as the neuron that fired also does not inhibit itself, and finally a larger amount of postsynaptic current is injected. The mean firing rates $\rho_{c}^j(t)$ of these Poisson process are however still defined by the softmax WTA rule given by (6). Consider now the the mean value of the postsynaptic current given by (8): 

\begin{equation}
\begin{aligned}
\mathbb{E}\left [ I_c^i(t)  \right ] &= \mathbb{E}\left [ \int_{0}^{t}\kappa(s) \cdot S_c^i(t-s)ds  \right ] \\
&= \int_{0}^{t}\kappa(s)\cdot \mathbb{E}\left [ S_c^i(t-s)ds  \right ] = \int_{0}^{t}\kappa(s)\cdot \rho(t-s)ds
\end{aligned}
\end{equation}

Here we've used the fact that the average spike-rate for a Poisson process is given by its rate-parameter $\lambda = \rho_{c}^i(t)$. In our network, we clamp the input nodes to the data and the feedforward dynamics are then allowed to operate. The average rates then settle down to their mean values. Hence after a long enough time, the average value of the mean-firing rates become constant. Under these steady state (s.s) conditions we get:

\begin{equation}
\mathbb{E}\left [ I_c^i(t)  \right ] = \int_{0}^{t}\kappa(s)\cdot \rho(t-s)ds \underset{s.s}= \rho \cdot \int_{0}^{t}\kappa(s)ds = \rho
\end{equation}

Note that in (40), $\rho = \underset{t \rightarrow \infty}\lim \rho(t)$ denotes the steady-state average firing rate. We have also used the the aforementioned normalization of $\kappa(t)$ to derive (37) from (36). We note that the settling time to reach this steady-state depends on how rapidly the kernel $\kappa(t)$ decays. In this instance, as we have used a double-exponential window, the steady-state will be achieved in approximately $5 \cdot \tau_s$ seconds (here $\tau_s$ is assumed to be larger than $\tau_f$, refer (7)). Further note that (40) deals with the expected value. However, we are also interested in quantifying the noisiness of this stochastic rate-coding scheme. We can quantify the average amount of fluctuation around the desired mean value by computing the steady state variance:

\begin{equation}
var\left ( I_c^i(t) \right ) \underset{s.s}= \rho \int_{0}^{t} \kappa^2(s)ds
\end{equation}

We can get a lower variance by scaling up the mean-firing rate to $\lambda_0 \cdot \rho_c^i(t)$ and multiplying the kernel $\kappa(t)$ by $1/\lambda_0$. This leads to the variance given by (41) to scale down by a factor of $\lambda_0$. For a derivation of (41) and the subsequent scaling result, we refer the reader to the end of this section. Intuitively, the result means that we can get a better estimate for the desired value by firing more often but injecting a smaller amount of current. We subsequently don't discuss the variance of our estimates, but note that the variance can be made arbitrarily small by choosing a large enough $\lambda_0$.
In summary, we have rate-coded the probabilities encoded by the WTA circuits and have $I_c^i(t) \approx \rho_{c}(t)$ with arbitrary precision.

Having described the dynamics of rate-coding in the soft spiking regime, we proceed to demonstrate that in this regime the network is implementing the feedforward pass of belief propagation on the variational posterior. We wish to show that $\rho_{c}^i(t) = q_f(z_c = i)$. We proceed to prove this inductively. Assume that $\rho_{r}^i(t) = q_f(z_{r} = i) \quad \forall r \in ch(c)$. By substituting this along with $\phi_r^{ij} = q(z_c=i|z_r=j)$ and $I_c^i(t) \approx \rho_{c}(t)$ in (16) we get:

\begin{equation}
d_{c,r}^i(t) = \log \left ( \sum_{i_r} \left ( q(z_c = i|z_{ch(c)^r} = i_r)q_f(z_{ch(c)^r} = i_r) \right ) \right )
\end{equation}

Note again that in (42) we first have current from all the children nodes adding up, noting as before that the marginals are simply the EPSP values and the conditional probabilities are the synaptic weights. This is then followed by the aforementioned logarithmic compression. The currents from all the dendrites sum up additively in the soma of the neuron according (15) to yield:

\begin{equation}
u_c^i(t) = \sum_{r=1}^{R}\left (\log \left ( \sum_{i_r} \left ( q(z_c = i|z_{ch(c)^r} = i_r)q_f(z_{ch(c)^r} = i_r) \right ) \right )  \right )
\end{equation}

Note that (43) is simply (38) in the log domain. Then by noting that the WTA exponentiates the membrane potential as per (6), we finally conclude that the soft WTA network is indeed performing feedforward message passing on the variational posterior. 

\subsection*{Variance of the rate-coding scheme}

In this subsection we will derive the expression for the variance of the rate-coded probabilities given by (41). To do this, we will use a limit representation of the Dirac delta function. Let $\Lambda_{\epsilon}(t)$ denote the rectangular function, with $\Lambda_{\epsilon}(t) = 1 \quad  \forall t \in (0,\epsilon)$ and $0$ otherwise. Then we have, $\delta_D(t) = \underset{t \rightarrow \infty}\lim \frac{1}{\epsilon}\Lambda_{\epsilon}(t)$, where $\delta_D(t)$ is the Dirac delta function, as noted before in Section 3.1. The spike-train $S_c^i(t)$ is a sum of Dirac delta functions, which we can approximate using $\Lambda_{\epsilon}(t)$. In an interval of length $ds$, $S_c^i(t)$ is then a binomial random variable that takes value $\frac{1}{ds}$ with probability $\lambda_0 \rho_c^i(t)ds$. Here $\lambda_0$ is a scaling factor, as mentioned in Section 4.1, we also rescale the current kernel as $\frac{\kappa(t)}{\lambda_0}$. We then have $\mathbb{E}\left [ S_c^i(t)ds  \right ] = \frac{1}{ds}ds\lambda_0\rho_{c}^i(t)ds = \lambda_0 \rho_{c}^i(t)ds$. Note that the mean current given by (39) and (40) are invariant under this rescaling:

\begin{equation}
\begin{aligned}
\mathbb{E}\left [ I_c^i(t)  \right ] &= \int_{0}^{t}\frac{\kappa(s)}{\lambda_0} \cdot \mathbb{E}\left [ S_c^i(t-s)ds  \right ] \\
&= \int_{0}^{t}\frac{\kappa(s)}{\lambda_0} \cdot \lambda_0 \cdot \rho(t-s)ds = \int_{0}^{t}\kappa(s) \cdot \rho(t-s)ds \underset{s.s}= \rho_{c}^i
\end{aligned}
\end{equation}

We can now these results to derive the variance under steady state conditions as:

\begin{equation}
\begin{aligned}
var\left ( I_c^i(t) \right ) &\underset{s.s}= \mathbb{E}\left [ (I_c^i(t))^2  \right ] - \mathbb{E}\left [ I_c^i(t)  \right ]^2 = \mathbb{E}\left [ (I_c^i(t))^2  \right ] - (\rho_c^i)^2 \\
&= \mathbb{E}\left [ \int_{0}^{t}\int_{0}^{t}\kappa(s_1) \cdot \kappa(s_2) \cdot S_c^i(t-s_1) \cdot S_c^i(t-s_2) ds_1ds_2  \right ] -(\rho_c^i)^2 \\
&= \int_{0}^{t}\int_{0}^{t}\kappa(s_1) \cdot \kappa(s_2) \cdot \mathbb{E}\left [  S_c^i(t-s_1) \cdot S_c^i(t-s_2) \right ] ds_1ds_2 -(\rho_c^i)^2
\end{aligned}
\end{equation}

Now, for $s_1 \neq s_2$ we have $\mathbb{E}\left [  S_c^i(t-s_1) \cdot S_c^i(t-s_2) \right ] = (\lambda_0 \cdot \rho_c^i(t-s_1)) \cdot (\lambda_0 \cdot \rho_c^i(t-s_2)) = (\lambda_0)^2 \cdot (\rho_c^i)^2$. And for $s_1 = s_2$ we have $\mathbb{E}\left [  S_c^i(t-s_1) \cdot S_c^i(t-s_2) \right ] = \mathbb{E}\left [  S_c^i(t-s_1) \cdot S_c^i(t-s_1) \right ] = \mathbb{E}\left [  S_c^i(t-s_1) \right ] = \frac{1}{ds} \cdot \frac{1}{ds} \cdot \lambda_0 \cdot \rho_c^i(t-s_1) ds = \frac{\lambda_0 \cdot \rho_c^i}{ds}$. This then yields:

\begin{equation}
\begin{aligned}
var\left ( I_c^i(t) \right ) &\underset{s.s}= \int_{0}^{t}\int_{0}^{t}\frac{\kappa(s_1)}{\lambda_0} \cdot \frac{\kappa(s_2)}{\lambda_0} \cdot \mathbb{E}\left [  S_c^i(t-s_1) \cdot S_c^i(t-s_2) \right ] ds_1ds_2 -(\rho_c^i)^2 \\
&= \left ( \int_{s_1 = s_2} + \int_{s_1 \neq s_2} \right )\frac{\kappa(s_1)}{\lambda_0} \cdot \frac{\kappa(s_2)}{\lambda_0} \cdot \mathbb{E}\left [  S_c^i(t-s_1) \cdot S_c^i(t-s_2) \right ] ds_1ds_2 - (\rho_c^i)^2 \\
&= \int_{0}^{t}\frac{\lambda_0 \cdot \rho_c^i}{ds_1} \cdot \frac{\kappa^2(s_1)}{\lambda_0^2}ds_1ds_1 + \int_{0}^{t}\int_{0}^{t}\frac{\kappa(s_1)}{\lambda_0} \cdot \frac{\kappa(s_2)}{\lambda_0} \cdot (\lambda_0)^2 \cdot (\rho_c^i)^2 ds_1ds_2 - (\rho_c^i)^2 \\
&= \frac{\rho_c^i}{\lambda_0} \cdot \int_{0}^{t} \kappa^2(s_1)ds_1 + (\rho_c^i)^2 \cdot \int_{0}^{t}\kappa(s_1) ds_1 \cdot \int_{0}^{t} \kappa(s_2)ds_2 -(\rho_c^i)^2 \\
&= \frac{\rho_c^i}{\lambda_0} \cdot \int_{0}^{t} \kappa^2(s_1)ds_1 + (\rho_c^i)^2 \cdot 1 \cdot 1 -(\rho_c^i)^2 = \frac{\rho_c^i}{\lambda_0} \cdot \int_{0}^{t} \kappa^2(s_1)ds_1
\end{aligned}
\end{equation}

Note that we have also used the fact that the line $s_1 = s_2$ has zero area. Also note that (41) is retrieved by setting $\lambda_0 = 1$ in (46).

\subsection{Learning via message passing}

In Section 3.2, the gradients with respect to the Free Energy as given by (24) and (25) were estimated via corresponding Monte-Carlo estimates given by (26) and (27). In this section we take a different approach by noting that (24) and (25) take the form of expectations with respect to the variational posterior. We proceed to show how these expectations can be estimated in terms of the marginals and conditional probabilities from the variational posterior, thus allowing us to approximately estimate them from the values of the firing rates and synaptic weights in this soft spiking regime. First, for (24) we have:

\begin{equation}
\begin{aligned}
-\nabla_{\theta_c^{ij}}F(x) &= \mathbb{E}_{Q}\left [ \frac{1}{\theta_c^{ij}}\delta(z_c=i) \cdot \delta(z_{pa(c)}=j) - \frac{1}{Z_{c}^{'j}}\delta(z_{pa(c)}=j) \right ] \\
&=\sum_{\alpha, \beta} \left ( \frac{1}{\theta_c^{ij}}\delta(z_c=i) \cdot \delta(z_{pa(c)}=j) - \frac{1}{Z_{c}^{'j}}\delta(z_{pa(c)}=j) \right ) \cdot q(z_c = \alpha, z_{pa(c)} = \beta) \\
&= \frac{1}{\theta_c^{ij}}q(z_c = i, z_{pa(c)} = j) - \frac{1}{Z_{c}^{'j}}q(z_{pa(c)} = j)
\end{aligned}
\end{equation}

We next proceed to approximate (25), which is more difficult due to the presence of the error term $e(x,h)$ inside the expectation. Computing the expression in (25) exactly requires clamping $z_c$ to $i$ and $z_{pa(c)}$ to $j$, allowing the network dynamics to settle to the estimates for the marginals, and to then use these estimates to compute the desired expression. This process then needs to be repeated \emph{sequentially} for every synapse in the network. It seems unlikely that such clamping happens in the brain. We instead propose an approximation to the gradient:

\begin{equation}
\begin{aligned}
-\nabla_{\phi_r^{ij}}F(x) &\approx \mathbb{E}_{Q}\left [e(x,h) \right )] \cdot \mathbb{E}_{Q}\left [\frac{1}{\phi_r^{ij}}\delta(z_c=i) \cdot \delta(z_r=j) - \frac{1}{Z_r^j}\delta(z_r=j) \right ] \\
&= \mathbb{E}_{Q}\left [e(x,h) \right )] \cdot \left [\frac{1}{\phi_r^{ij}}q(z_c=i,z_r=j) - \frac{1}{Z_r^j}q(z_r=j) \right ]
\end{aligned}
\end{equation}

Thus we are using an approximation of the form $\mathbb{E} \left [A \cdot B \right] \approx \mathbb{E} \left [A \right] \cdot \mathbb{E} \left [B \right]$. The error in such an approximation is the covariance of the two random variables because $\mathbb{E} \left [(A - \mathbb{E}(A)) \cdot (B - \mathbb{E}(B)) \right] = \mathbb{E} \left [A \cdot B \right] - \mathbb{E} \left [A \right] \cdot \mathbb{E} \left [B \right]$. We thus note that this is a biased estimate for the gradient, unlike the unbiased estimate obtained via Monte-Carlo sampling in Section 3.2.

As we had done in Section 3.2, we proceed to rescale (47) and (48). We again assume that the normalization constants were all equal to $1$ to begin with. Also, we note that $\mathbb{E}_{Q}\left [e(x,h) \right )] = D_{KL}\left ( Q_{\phi}(h|x) || P_{\theta}(h|x) \right ) > 0$. Hence $\mathbb{E}_{Q}\left [e(x,h) \right )]$ is a non-negative number and we drop this term from the expression for the rescaled gradients. This is a reasonable thing to do here because by using rescaled the gradients, we are discarding the magnitude of the gradient (and are only using the direction) and hence any non-negative quantity can be also discarded in this process. We rescale (47) and (48) respectively by $\theta_c^{ij}$ and $\phi_r^{ij}$ to obtain the weight update rules:

\begin{align}
\Delta \theta_c^{ij} &= q(z_c=i,z_{pa(c)}=j) - \theta_c^{ij} \cdot q(z_{pa(c)}=j)\\
\Delta \phi_r^{ij} &= q(z_c=i,z_r=j) - \phi_r^{ij} \cdot q(z_r=j)
\end{align}

Further, the normalization constants don't change after the weight updates given by (49) and (50):

\begin{align}
&\begin{aligned}
\Delta Z_c^{'j} &= \sum_{i=1}^{K}\Delta \theta_c^{ij} = \sum_{i=1}^{K} \left ( q(z_c=i,z_{pa(c)}=j) - \theta_c^{ij} \cdot q(z_{pa(c)}=j) \right ) \\
&= \sum_{i=1}^{K} q(z_c=i,z_{pa(c)}=j) - \sum_{i=1}^{K} \theta_c^{ij} \cdot q(z_{pa(c)}=j) \\
&= q(z_{pa(c)}=j) - 1 \cdot q(z_{pa(c)}=j = 0 
\end{aligned}\\
&\begin{aligned}
\Delta Z_r^{j} &= \sum_{i=1}^{K}\Delta \phi_r^{ij} = \sum_{i=1}^{K} \left ( q(z_c=i,z_r=j) - \phi_r^{ij} \cdot q(z_r=j) \right ) \\
&= \sum_{i=1}^{K} q(z_c=i,z_r=j)- \sum_{i=1}^{K} \phi_r^{ij} \cdot q(z_r=j) \\
&= q(z_r=j) - 1 \cdot q(z_r=j) = 0
\end{aligned}
\end{align}

To express the terms in (49) and (50) in terms of the neural dynamics, we first note from the discussion on rate-coding that $q(z_{pa(c)} = j) \approx \rho_{pa(c)}^j(t)$. To compute the joint distribution $q(z_c = i, z_{pa(c)} = j)$, we follow a marginalisation strategy similar to (38), but this time we keep both $z_c$ and $z_{pa(c)}$ fixed and marginalized over the rest of the variables. Let the $m_{c,pa(c)}$ and $m_{pa(c),c}$ respectively denote the set of messages from $c$ to $pa(c)$, and from $pa(c)$ to $c$. For brevity, let's call them $a_c(t)$ and $b_c(t)$ respectively. Then we have $a_c^j(t) = \sum_{i=1}^{K} \phi_c^{ji} \cdot \rho_c^i(t)$, and $b_c^i(t) = \sum_{j=1}^{K} \phi_c^{ji} \cdot \rho_{pa(c)}^j(t)$, which is precisely the net incident input on the dendrite on which these synapses are incident. We then have: 

\begin{align}
\begin{split}
q(z_{pa(c)} = j, z_c = i) &\propto  \frac{q(z_{pa(c)} = j|z_{c} = i) \cdot q(z_{c} = i) \cdot q(z_{pa(c)} = j)}{m_{c,pa(c)}^j \cdot m_{pa(c),c}^i} \\
&\approx 
\frac{\phi_c^{ji} \cdot \rho_c^i(t) \cdot \rho_{pa(c)}^j(t)}{a_c^j(t) \cdot b_c^i(t)}
\end{split}\\
\begin{split}
q(z_c=i,z_r=j) &\propto  \frac{q(z_{c} = i|z_{r} = j) \cdot q(z_{c} = i) \cdot q(z_{r} = j)}{m_{c,r}^j \cdot m_{r,c}^i} \\
&\approx 
\frac{\phi_r^{ij} \cdot \rho_c^i(t) \cdot \rho_{r}^j(t)}{a_r^i(t) \cdot b_r^j(t)} 
\end{split}
\end{align}

Thus we see that the expression for the gradient involves the net incident input on the dendrite on which the synapse is located, as well as the firing rates of the presynaptic and postsynaptic neurons. Furthermore, these rate-based gradients can be estimated using spike-trains. First, we have $\mathbb{E}\left [ S_{c}^i(t) \right ] = \rho_{c}^i(t)$, and thus $S_{c}^i(t)$ is an unbiased estimate for $\rho_{c}^i(t)$. Furthermore, we also have:

\begin{equation}
\begin{aligned}
\mathbb{E}_Q\left [S_c^i(t) \cdot \int_{0}^{t}S_{pa(c)}^j(t-s) \cdot W(s)ds \right ] &= \mathbb{E}\left [S_c^i(t) \right ] \cdot \int_{0}^{t}\mathbb{E}\left [ S_{pa(c)}^j(t-s) \right ] \cdot W(s)ds\\
= \rho_c^i(t) \cdot \int_{0}^{t}\rho_{pa(c)}^j(t-s)\cdot W(s)ds
&\underset{t \rightarrow \infty}\rightarrow 
\rho_c^i(t) \cdot \rho_{pa(c)}^j(t) \cdot \int_{0}^{\infty}W(s)ds
= \rho_c^i(t) \cdot \rho_{pa(c)}^j(t)  
\end{aligned}
\end{equation}

Note that here we have used the fact that $S_c^i(t)$, $S_{pa(c)}^j(t)$ and $S_r^j(t)$ are all indepedent Poisson processes, and thus the expectation of their product factors into the product of their expectations. Thus we see that we can use $S_c^i(t) \cdot \int_{0}^{t}S_{pa(c)}^j(t-s) \cdot W(s)ds$ as an unbiased estimate for $\rho_c^i(t) \cdot \rho_{pa(c)}^j(t)$. As with the hard WTA case, we choose our learning rules as $\xi_r^{ij}(t) = \frac{1}{N_r^j(t)}$ and $\xi_c^{'ij}(t) = \frac{1}{N_c^i(t)}$ to satisfy the conditions of the Robbins-Monro algorithm \cite{borkar2013stochastic}. We also note that there is a bias-variance tradeoff between learning in the hard and soft case. The estimates for the gradients obtained in the hard case are unbiased, but have higher variance, as we are working with a trivial Monte-Carlo estimate. The inherent stochasticity of the Poisson firing further increases the variance. On the other hand, the estimates for the gradients in the soft case are biased, but have lower variance, as we are explicitly approximating the expressions for the expectations in the gradients and in this case the variance only comes from the Poisson firing.

\bibliography{Bibliography}

\begin{thebibliography}{10}

\bibitem{doya2007bayesian}
K.~Doya, {\em Bayesian brain: Probabilistic approaches to neural coding}.
\newblock MIT press, 2007.

\bibitem{knill2004bayesian}
D.~C. Knill and A.~Pouget, ``The bayesian brain: the role of uncertainty in
  neural coding and computation,'' {\em TRENDS in Neurosciences}, vol.~27,
  no.~12, pp.~712--719, 2004.

\bibitem{blei2017variational}
D.~M. Blei, A.~Kucukelbir, and J.~D. McAuliffe, ``Variational inference: A
  review for statisticians,'' {\em Journal of the American Statistical
  Association}, vol.~112, no.~518, pp.~859--877, 2017.

\bibitem{zhang2018advances}
C.~Zhang, J.~Butepage, H.~Kjellstrom, and S.~Mandt, ``Advances in variational
  inference,'' {\em IEEE transactions on pattern analysis and machine
  intelligence}, 2018.

\bibitem{shimazaki2019principles}
H.~Shimazaki, ``The principles of adaptation in organisms and machines i:
  machine learning, information theory, and thermodynamics,'' {\em arXiv
  preprint arXiv:1902.11233}, 2019.

\bibitem{friston2010free}
K.~Friston, ``The free-energy principle: a unified brain theory?,'' {\em Nature
  reviews neuroscience}, vol.~11, no.~2, p.~127, 2010.

\bibitem{pecevski2011probabilistic}
D.~Pecevski, L.~Buesing, and W.~Maass, ``Probabilistic inference in general
  graphical models through sampling in stochastic networks of spiking
  neurons,'' {\em PLoS computational biology}, vol.~7, no.~12, p.~e1002294,
  2011.

\bibitem{nessler2013bayesian}
B.~Nessler, M.~Pfeiffer, L.~Buesing, and W.~Maass, ``Bayesian computation
  emerges in generic cortical microcircuits through spike-timing-dependent
  plasticity,'' {\em PLoS computational biology}, vol.~9, no.~4, p.~e1003037,
  2013.

\bibitem{habenschuss2012homeostatic}
S.~Habenschuss, J.~Bill, and B.~Nessler, ``Homeostatic plasticity in bayesian
  spiking networks as expectation maximization with posterior constraints,'' in
  {\em Advances in Neural Information Processing Systems}, pp.~773--781, 2012.

\bibitem{kappel2014stdp}
D.~Kappel, B.~Nessler, and W.~Maass, ``Stdp installs in winner-take-all
  circuits an online approximation to hidden markov model learning,'' {\em PLoS
  computational biology}, vol.~10, no.~3, p.~e1003511, 2014.

\bibitem{pecevski2016learning}
D.~Pecevski and W.~Maass, ``Learning probabilistic inference through
  spike-timing-dependent plasticity,'' {\em eneuro}, vol.~3, no.~2, 2016.

\bibitem{yu2018winner}
Z.~Yu, Y.~Tian, T.~Huang, and J.~K. Liu, ``Winner-take-all as basic
  probabilistic inference unit of neuronal circuits,'' {\em arXiv preprint
  arXiv:1808.00675}, 2018.

\bibitem{guo2017hierarchical}
S.~Guo, Z.~Yu, F.~Deng, X.~Hu, and F.~Chen, ``Hierarchical bayesian inference
  and learning in spiking neural networks,'' {\em IEEE transactions on
  cybernetics}, no.~99, pp.~1--13, 2017.

\bibitem{cort2013conditional}
B.~Cort and B.~Anderson, ``Conditional probability modulates visual search
  efficiency,'' {\em Frontiers in human neuroscience}, vol.~7, p.~683, 2013.

\bibitem{schmitz2014components}
F.~Schmitz and A.~Voss, ``Components of task switching: A closer look at task
  switching and cue switching,'' {\em Acta psychologica}, vol.~151,
  pp.~184--196, 2014.

\bibitem{wainwright2008graphical}
M.~J. Wainwright, M.~I. Jordan, {\em et~al.}, ``Graphical models, exponential
  families, and variational inference,'' {\em Foundations and
  Trends{\textregistered} in Machine Learning}, vol.~1, no.~1--2, pp.~1--305,
  2008.

\bibitem{hoffman2013stochastic}
M.~D. Hoffman, D.~M. Blei, C.~Wang, and J.~Paisley, ``Stochastic variational
  inference,'' {\em The Journal of Machine Learning Research}, vol.~14, no.~1,
  pp.~1303--1347, 2013.

\bibitem{mnih2014neural}
A.~Mnih and K.~Gregor, ``Neural variational inference and learning in belief
  networks,'' {\em arXiv preprint arXiv:1402.0030}, 2014.

\bibitem{dayan1995helmholtz}
P.~Dayan, G.~E. Hinton, R.~M. Neal, and R.~S. Zemel, ``The helmholtz machine,''
  {\em Neural computation}, vol.~7, no.~5, pp.~889--904, 1995.

\bibitem{sountsov2015spiking}
P.~Sountsov and P.~Miller, ``Spiking neuron network helmholtz machine,'' {\em
  Frontiers in computational neuroscience}, vol.~9, p.~46, 2015.

\bibitem{shin1991pi}
Y.~Shin and J.~Ghosh, ``The pi-sigma network: An efficient higher-order neural
  network for pattern classification and function approximation,'' in {\em
  IJCNN-91-Seattle International Joint Conference on Neural Networks}, vol.~1,
  pp.~13--18, IEEE, 1991.

\bibitem{wu2018and}
L.~Wu, Y.~Wang, X.~Li, and J.~Gao, ``What-and-where to match: Deep spatially
  multiplicative integration networks for person re-identification,'' {\em
  Pattern Recognition}, vol.~76, pp.~727--738, 2018.

\bibitem{trask2018neural}
A.~Trask, F.~Hill, S.~E. Reed, J.~Rae, C.~Dyer, and P.~Blunsom, ``Neural
  arithmetic logic units,'' in {\em Advances in Neural Information Processing
  Systems}, pp.~8035--8044, 2018.

\bibitem{schmitt2002complexity}
M.~Schmitt, ``On the complexity of computing and learning with multiplicative
  neural networks,'' {\em Neural Computation}, vol.~14, no.~2, pp.~241--301,
  2002.

\bibitem{yedidia2003understanding}
J.~S. Yedidia, W.~T. Freeman, and Y.~Weiss, ``Understanding belief propagation
  and its generalizations,'' {\em Exploring artificial intelligence in the new
  millennium}, vol.~8, pp.~236--239, 2003.

\bibitem{jones2012logarithmic}
P.~W. Jones and F.~Gabbiani, ``Logarithmic compression of sensory signals
  within the dendritic tree of a collision-sensitive neuron,'' {\em Journal of
  Neuroscience}, vol.~32, no.~14, pp.~4923--4934, 2012.

\bibitem{keil2015dendritic}
M.~S. Keil, ``Dendritic pooling of noisy threshold processes can explain many
  properties of a collision-sensitive visual neuron,'' {\em PLoS computational
  biology}, vol.~11, no.~10, p.~e1004479, 2015.

\bibitem{rao2005hierarchical}
R.~P. Rao, ``Hierarchical bayesian inference in networks of spiking neurons,''
  in {\em Advances in neural information processing systems}, pp.~1113--1120,
  2005.

\bibitem{hawkins2016neurons}
J.~Hawkins and S.~Ahmad, ``Why neurons have thousands of synapses, a theory of
  sequence memory in neocortex,'' {\em Frontiers in neural circuits}, vol.~10,
  p.~23, 2016.

\bibitem{lee2003hierarchical}
T.~S. Lee and D.~Mumford, ``Hierarchical bayesian inference in the visual
  cortex,'' {\em JOSA A}, vol.~20, no.~7, pp.~1434--1448, 2003.

\bibitem{sacramento2018dendritic}
J.~Sacramento, R.~P. Costa, Y.~Bengio, and W.~Senn, ``Dendritic cortical
  microcircuits approximate the backpropagation algorithm,'' in {\em Advances
  in Neural Information Processing Systems}, pp.~8721--8732, 2018.

\bibitem{borkar2013stochastic}
V.~S. Borkar, ``Stochastic approximation,'' {\em Resonance}, vol.~18, no.~12,
  pp.~1086--1094, 2013.

\bibitem{burbank2015mirrored}
K.~S. Burbank, ``Mirrored stdp implements autoencoder learning in a network of
  spiking neurons,'' {\em PLoS computational biology}, vol.~11, no.~12,
  p.~e1004566, 2015.

\bibitem{kolen1994backpropagation}
J.~F. Kolen and J.~B. Pollack, ``Backpropagation without weight transport,'' in
  {\em Proceedings of 1994 IEEE International Conference on Neural Networks
  (ICNN'94)}, vol.~3, pp.~1375--1380, IEEE, 1994.

\bibitem{akrout2019deep}
M.~Akrout, C.~Wilson, P.~C. Humphreys, T.~P. Lillicrap, and D.~B. Tweed, ``Deep
  learning without weight transport.,'' {\em CoRR, abs/1904.05391}, 2019.

\bibitem{wainwright2003tree}
M.~J. Wainwright, T.~S. Jaakkola, and A.~S. Willsky, ``Tree-based
  reparameterization framework for analysis of sum-product and related
  algorithms,'' {\em IEEE Transactions on information theory}, vol.~49, no.~5,
  pp.~1120--1146, 2003.

\bibitem{raju2016inference}
R.~V. Raju and Z.~Pitkow, ``Inference by reparameterization in neural
  population codes,'' in {\em Advances in Neural Information Processing
  Systems}, pp.~2029--2037, 2016.

\bibitem{parr2019neuronal}
T.~Parr, D.~Markovic, S.~J. Kiebel, and K.~J. Friston, ``Neuronal message
  passing using mean-field, bethe, and marginal approximations,'' {\em
  Scientific reports}, vol.~9, no.~1, p.~1889, 2019.

\end{thebibliography}
\bibliographystyle{ieeetr}

\end{document}